\documentclass[manuscript,screen,nonacm,sigconf]{acmart}

\AtBeginDocument{%
  \providecommand\BibTeX{{%
    \normalfont B\kern-0.5em{\scshape i\kern-0.25em b}\kern-0.8em\TeX}}}

\setcopyright{none}
\copyrightyear{2024}
\acmYear{2024}
\acmDOI{XXXXXXX.XXXXXXX}

\acmConference[DAC '24]{}{June 23--27,
  2024}{San Francisco, CA}
\usepackage{comment}
\renewenvironment{comment}{}{}
\usepackage{soul}
\usepackage{enumitem}
\usepackage{multirow}
\usepackage{makecell}
\usepackage{subcaption}

\excludecomment{daccomment}

\acmSubmissionID{1518}

\begin{document}

\title{A Deep Learning Accelerator is \textbf{NOT} All You Need}
\title{Performance Analysis of CPU-GPU Servers for Computer Vision DNNs}
\title{Performance of CPU-GPU Servers for Computer Vision DNNs}
\title{Beyond Inference: A Comprehensive Performance Analysis of DNN Server Overheads for Computer Vision Applications}

\title{The Overlooked Performance Bottlenecks in DNN-Based Computer Vision Serving Systems}
\title{Beyond Inference: Performance Analysis of DNN Server Overheads for Computer Vision}


% \author{Ahmed F. AbouElhamayed$^*$, Susanne Balle$^\mathsection$, Deshanand Singh$^\mathsection$, Mohamed S. Abdelfattah$^*$}
% \affiliation{
% \institution{Cornell University$^*$, Intel$^\mathsection$}
% }

\author{\fontsize{10.5}{11}\selectfont Ahmed F. AbouElhamayed}
\email{afa55@cornell.edu}
\orcid{0000-0001-6381-2936}
\affiliation{%
  \institution{Cornell University}
  \state{New York}
  \country{USA}
}
\vspace{-5mm} % Adjust the negative value as needed
\author{\fontsize{10.5}{11}\selectfont Susanne Balle}
\email{susanne.balle@intel.com}
\orcid{}
\affiliation{%
  \institution{Intel}
  \state{Massachusetts}
  \country{USA}
}
\author{\fontsize{10.5}{11}\selectfont Deshanand Singh}
\email{deshanand.singh@intel.com}
\orcid{0009-0003-4968-4343}
\affiliation{%
  \institution{Intel}
  \state{Ontario}
  \country{Canada}
}

\author{\fontsize{10.5}{11}\selectfont Mohamed S. Abdelfattah}
\email{mohamed@cornell.edu}
% \orcid{0000-0001-6381-2936}
\orcid{0000-0002-4568-8932}
\affiliation{%
  \institution{Cornell University}
  \state{New York}
  \country{USA}
}

% Adding comments
\newif\ifcomment
\commenttrue

\ifcomment
\newcommand{\ahmed}[1]{\sethlcolor{cyan}\hl{[Ahmed: #1]}}
\newcommand{\moh}[1]{\sethlcolor{green}\hl{[Moh: #1]}}
\newcommand{\draftfigure}[1]{\sethlcolor{blue}\hl{[Figure: #1]}}
\newcommand{\important}[1]{\sethlcolor{red}\hl{[IMP: #1]}}
\newcommand{\m}[1]{\sethlcolor{mohamed_colour}\hl{[Moh: #1]}}
\else
\newcommand{\ahmed}[1]{}
\newcommand{\moh}[1]{}
\newcommand{\important}[1]{}
\newcommand{\m}[1]{}
\fi

\renewcommand{\comment}[1]{}
\definecolor{mohamed_colour}{RGB}{255, 204, 205}

\renewcommand{\shortauthors}{AbouElhamayed, et al.}

\begin{abstract}
Deep neural network (DNN) inference has become an important part of many data-center workloads.
This has prompted focused efforts to design ever-faster deep learning accelerators such as GPUs and TPUs.
However, an end-to-end DNN-based vision application contains more than just DNN inference, including input decompression, resizing, sampling, normalization, and data transfer.
In this paper, we perform a thorough evaluation of computer vision inference requests performed on a throughput-optimized serving system.
We quantify the performance impact of server overheads such as data movement, preprocessing, and message brokers between two DNNs producing outputs at different rates.
Our empirical analysis encompasses many computer vision tasks including image classification, segmentation, detection, depth-estimation, and more complex processing pipelines with multiple DNNs.
%Furthermore, we test multiple systems with different CPU-GPU configurations.
%
Our results consistently demonstrate that end-to-end application performance can easily be dominated by data processing and data movement functions (up to 56\% of end-to-end latency in a medium-sized image, and $\sim$ 80\% impact on system throughput in a large image), even though these functions have been conventionally overlooked in deep learning system design.
Our work identifies important performance bottlenecks in different application scenarios, achieves 2.25$\times$ better throughput compared to prior work, and paves the way for more holistic deep learning system design.
\end{abstract}

\comment{
\begin{CCSXML}
<ccs2012>
   <concept>
       <concept_id>10010147.10010178</concept_id>
       <concept_desc>Computing methodologies~Artificial intelligence</concept_desc>
       <concept_significance>500</concept_significance>
       </concept>
   <concept>
       <concept_id>10010520</concept_id>
       <concept_desc>Computer systems organization</concept_desc>
       <concept_significance>500</concept_significance>
       </concept>
 </ccs2012>
\end{CCSXML}

\ccsdesc[500]{Computing methodologies~Artificial intelligence}
\ccsdesc[500]{Computer systems organization}

\keywords{Deep Learning Inference, Machine Learning Systems, Benchmark, Preprocessing}
}


% \received{11 November 2023}
% \received[revised]{12 March 2024}

\maketitle

\vspace{-.3cm}

\section{Introduction}
\label{sec:introduction}

% intro to CV serving systems and a motivation for performance
Artificial Intelligence (AI) has quickly proliferated different aspects of computing as exemplified by the rise of large language models (LLMs) within chat bots like ChatGPT---the fastest growing consumer application in history~\cite{llmlargestgrowing}. 
Such deep neural networks (DNNs) need to handle millions or billions of real-time customer queries on a daily basis, requiring fast, scalable, and efficient inference systems. 
This is equally true for computer vision applications.
For instance, social media platforms like Facebook process more than 10.5 billion photos per month \cite{fbphotos} using DNNs for person detection, automatic tagging, content classification, and style transfer. 
Furthermore, video broadcasting platforms such as YouTube use DNNs to automatically detect age-restricted content in addition to other tasks like auto-captioning~\cite{youtube_autodetect}.
Many of these computer vision tasks require fast and throughput-oriented inference servers to perform efficient DNN inference.
In addition, many of these tasks contain non-trivial preprocessing and postprocessing functions to compress and manipulate the DNN input data for both efficiency and compatibility with the DNN. These functions, referred to as \textit{DNN inference overheads}, are particularly common in applications involving image and video data which is large, high-dimensional, and comes in many different sizes, formats, and properties.

% optimizing DNN is not enough with examples from industry and academia on new deep learning hardware
To fulfill an inference request, a trained DNN is deployed on a web server.
These modern servers have begun to include dedicated hardware for inference, including Graphical Processing units (GPUs), and custom ASICs like Google's Tensor Processing Unit (TPU)~\cite{jouppi2023tpu}, Amazon's Inferentia~\cite{zheng2020optimizing}, Meta's MTIA~\cite{firoozshahian2023mtia}, in addition to academic deep learning accelerators~\cite{abdelfattah2018dla}.
These accelerators have been immensely successful at accelerating the massively-parallel matrix multiplications that are present in DNN inference, however, other data manipulation tasks such as decompression, resizing, decoding, and sampling often need to \textit{fall-back} to the host CPU for preprocessing.
For example, a video classification service receives the video in a compressed format like MPEG, decodes the video, samples a number of frames, then resizes and normalizes the resulting images into the format required by the DNN. 
The current emphasis on enhancing DNN performance while disregarding other tasks will eventually limit performance according to Amdahl's Law~\cite{amdahl1967validity}.
This motivates a deeper investigation into the performance bottlenecks and latency breakdowns of deep-learning vision tasks on modern server systems.

In this work, we quantify the performance impact of DNN inference overheads for computer vision applications running on state-of-the-art inference systems to better understand and optimize end-to-end system performance.
On an optimized serving system, we implement and profile a number of computer vision applications including image classification, segmentation, object detection, depth-estimation, and a multi-DNN application with message brokers.
In addition to highlighting the impact of data processing, we study the performance and energy efficiency of different server configurations, different data processing hardware options, and implications of multi-GPU systems.
Our work advances our understanding of DNN inference systems, quantifies performance bottlenecks, and advocates for holistic end-to-end DNN inference optimization. 
Our key contributions are:

\vspace{-.3cm}

\begin{enumerate}[label=\textbf{(\arabic*)}]
%[wide, labelwidth=!, labelindent=0pt,label=\textbf{(\arabic*)}]
%\itemsep0em
\item Quantify the latency, throughput, and energy efficiency implications of DNN inference overheads within deep-learning based computer vision applications. We find that non-DNN data processing can take up to 56\% of the time it takes to process a DNN inference request in a medium-sized image, even within a highly optimized inference server.
\item Evaluate different hardware systems for DNN serving systems, including CPU-GPU, GPU-only, and CPU-multi-GPU systems to investigate scaling. We draw many conclusions about optimized system setup and CPU-GPU ratios.
\item Investigate and optimize a computer vision pipeline with two DNN inference calls (detection then identification), demonstrating 2.25$\times$ higher throughput compared to prior work.
\end{enumerate}
%
%=================================================================
\section{Background}

\subsection{Inference Systems}

\begin{figure}[!t]
    \centering
    \includegraphics[width=0.8\linewidth]{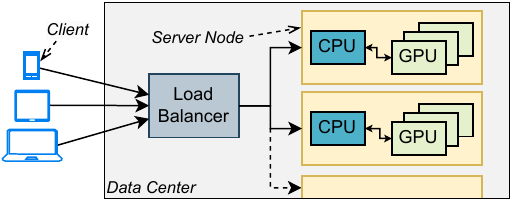}
        \vspace*{-3mm}
    \caption{Sample API system serving a DNN on a GPU.}
    % \vspace*{-10mm}
    \label{fig:api_system_components}
\end{figure}

% explain serving system and the application scenario that we are experimenting with
% One of the common ways to utilize DNNs within applications is through server deployment that is accessible on client devices (e.g. mobile phones or laptops) via an Application Programming Interface (API). 
% In this scenario, a \textit{load balancer} pools requests within the datacenter, then distributes them across the available processing servers as shown in Fig.~\ref{fig:api_system_components}.
% Each server must manage concurrent requests, necessitating multi-threading and concurrency control mechanisms. 
% We consider the case wherein the number of concurrent requests per server is capped at a certain value, guaranteed by the load balancer, then more servers are added if the number of requests increase beyond the system capacity.

A common approach to deploying DNNs in various applications involves server-based execution, accessible to client devices (such as mobile phones or laptops) through an Application Programming Interface (API). 
In this model, a \textit{load balancer} within the datacenter receives incoming requests and strategically distributes them among the available processing servers, as illustrated in Fig~\ref{fig:api_system_components}. 
Each server handles multiple concurrent requests, necessitating the implementation of multi-threading and concurrency control strategies. 
Our study focuses on a scenario where the load balancer imposes a cap on the number of concurrent requests each server can handle. 
In instances where incoming requests exceed the system’s predefined capacity, additional servers are added to maintain performance.

% make things more specific to ML and GPU-based servers
To efficiently process DNNs, GPUs or custom ASICS (e.g. TPUs) have become necessary in modern datacenters.
These devices are optimized for batch processing, thereby presenting a challenge to a server's typical individual request handling.
This is a key reason for creating specialized DNN inference serving software such as NVIDIA's Triton Inference Server (TrIS). 
Key to its operation is \textit{dynamic batching}, which aggregates incoming requests for batch processing often while ensuring bounded latency. 
Serving software provides many adjustable settings, including the maximum queuing latency, and maximum batch size.
%Furthermore, to ensure acceptable processing latency, there's often a preference to cap the batch size. 
%In response, software settings are made available to define this maximum batch size, ensuring that processing is never bogged down by overly large batches, irrespective of the volume of incoming requests.
Additionally, multiple \textit{instances}\footnote{Processes in case of CPU and CUDA Streams in case of GPU.} of the processing units can each handle requests independently, which in turn increases the number of requests the server can handle at a time. 

% talk about ways of measuring server effectiveness?
There are two key measures of server performance.
The first is throughput that quantifies the number of requests that can be processed per second.
The second is latency, to measure how long each request takes. 
Such servers typically produce a distribution of latencies as a result of differing arrival times, dynamic batches, and CPU load, therefore, average and tail latency are typically reported to represent the typical and worst-case server performance respectively. 

\begin{figure}[!h]
    \centering
    \includegraphics[width=1\linewidth]{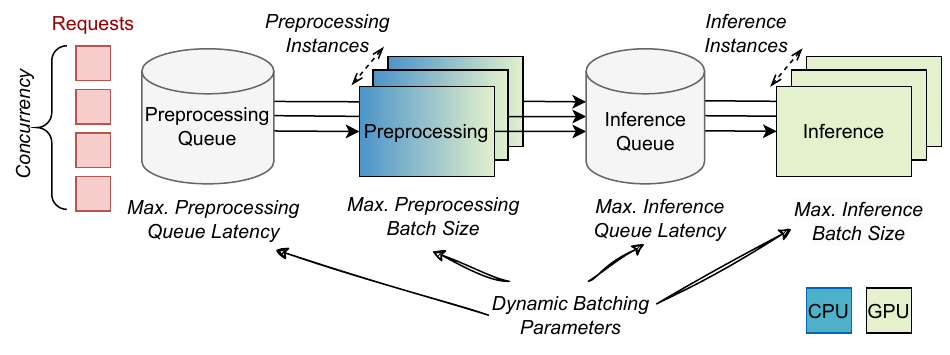}
    \vspace*{-8mm}
    \caption{A sample DNN application consisting of preprocessing and inference with annotated server parameters.}
    \vspace*{-2mm}
    \label{fig:server_parameters}
\end{figure}
\vspace{-.3cm}

\subsection{Application Scenarios}

% a description of our applications, including the fact that we have rate-matched preprocess-infer and different-rate detection-classification pipelines
A typical DNN inference pipeline is demonstrated in Fig.~\ref{fig:server_parameters}.
In addition to a DNN, this system contains a preprocessing stage which transforms user data to the correct size and format required by the DNN.
Typically, general preprocessing is handled by a server CPU, while the DNN inference is offloaded to an accelerator such as the GPU.
However, the increase in GPU speed and stagnation in CPU performance has necessitated accelerated preprocessing solutions.
This is why common preprocessing functions, especially those related to image and video data, are being accelerated on GPUs through software libraries like NVIDIA DALI~\cite{nvidihadali}.
The importance of preprocessing performance is further underscored by the inclusion of a dedicated hardware JPEG decoder specifically for DNN preprocessing on modern GPUs such as NVIDIA A100~\cite{nvidiaa100}. 
%This strongly motivates our in-depth study on preprocessing overheads in this paper.

Fig.~\ref{fig:server_parameters} demonstrates a two-stage application pipeline in which both the first and second stages are \textit{rate-matched}.
However, in other scenarios one input to stage 1 can produce multiple outputs that need to be processed by stage 2.
The example that we study further in Section~\ref{sec:multi-dnn} consists of a face detection pipeline followed by identification.
In this case, a single frame could contain multiple faces to be processed by the face identification stage.
In this case, typically a \textit{message broker}---such as Apache Kafka or Redis---is used between the producer and consumer to manage communication between the two processing stages.

\vspace{-.3cm}

\subsection{Software Configuration Impact}
\label{sec:software_impact}
\begin{figure}[t]
    \centering
    \includegraphics[width=1\linewidth, trim = 0 10mm 0 0]{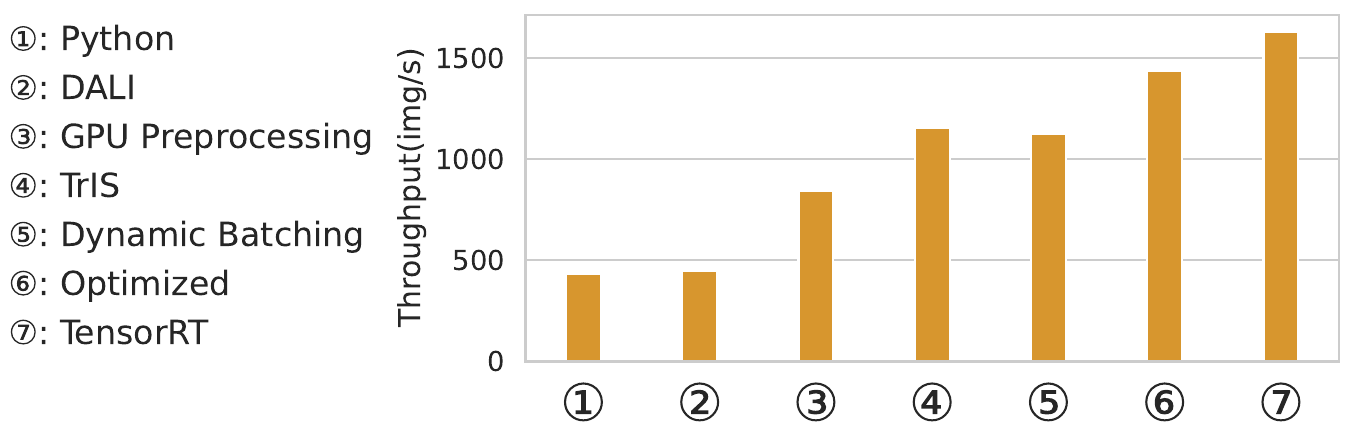}
    % \vspace*{-8mm}
    \caption{Evaluation of throughput across diverse system setups running the same Vision Transformer (ViT) model.}
    %\vspace*{-4mm}

    \label{fig:motivation-performance-on-different-systems}
\end{figure}
% Experiments: 256 - to be replaced with 382,266 - to be replaced with 383,267,268,269,270,271,272
%

To accurately quantify DNN inference overheads, we must do so under optimized server configurations.
Fig~\ref{fig:motivation-performance-on-different-systems} highlights the stark performance differences on the same hardware platform while applying different optimizations.
On a CPU-GPU system\footnote{Our setup is a dedicated server node with 13th Gen Intel(R) Core(TM) i9-13900K CPU and an NVIDIA GeForce RTX 4090 GPU.}, we measure the performance of an image classification pipeline with the Vision Transformer (ViT) base model.
First we start with the PyTorch model downloaded directly from HuggingFace and we run it without any serving software, just a Python loop that decompresses JPEG images one-by-one, followed by batched DNN inference, yielding $\sim$431 img/s. 
Next, throughput increases to $\sim$446 img/s when we use NVIDIA's DALI framework that enables batched image decompression. When enabling GPU for preprocessing which uses the NvJPEG library, the throughput increases to $\sim$842 img/s.
Employing TrIS (with the ONNX runtime) instead of PyTorch further improves performance thanks to its more optimized model execution and asynchronous processing, allowing an overlap between computation and data movement. 
Next, we enable TrIS dynamic batching instead of a fixed batch size to mimic a realistic server workload. 
Even though throughput drops slightly, the tail latency improves from 55~ms to 38~ms, providing a better quality of service. 
To optimize the server setup, we perform a quick search on its settings that include the number of preprocessing and inference processes, the maximum allowed batch size, and the concurrency per server.
This results in a $\sim$300~img/s throughput improvement, showing that server software parameters are critical to high performance.
Finally, we use TensorRT to enable model-level optimizations while compiling ViT, resulting in additional throughput improvement to more than 1600~img/s. 
Overall, there is more than 8$\times$ throughput improvement between the baseline PyTorch and optimized TrIS+TensorRT implementations. 
This highlights the importance of optimizing the server setup to measure realistic and representative system performance---we use this optimized setup for all our results in this paper.

\begin{figure}[!t]
    \centering    
    \includegraphics[width=1\linewidth]{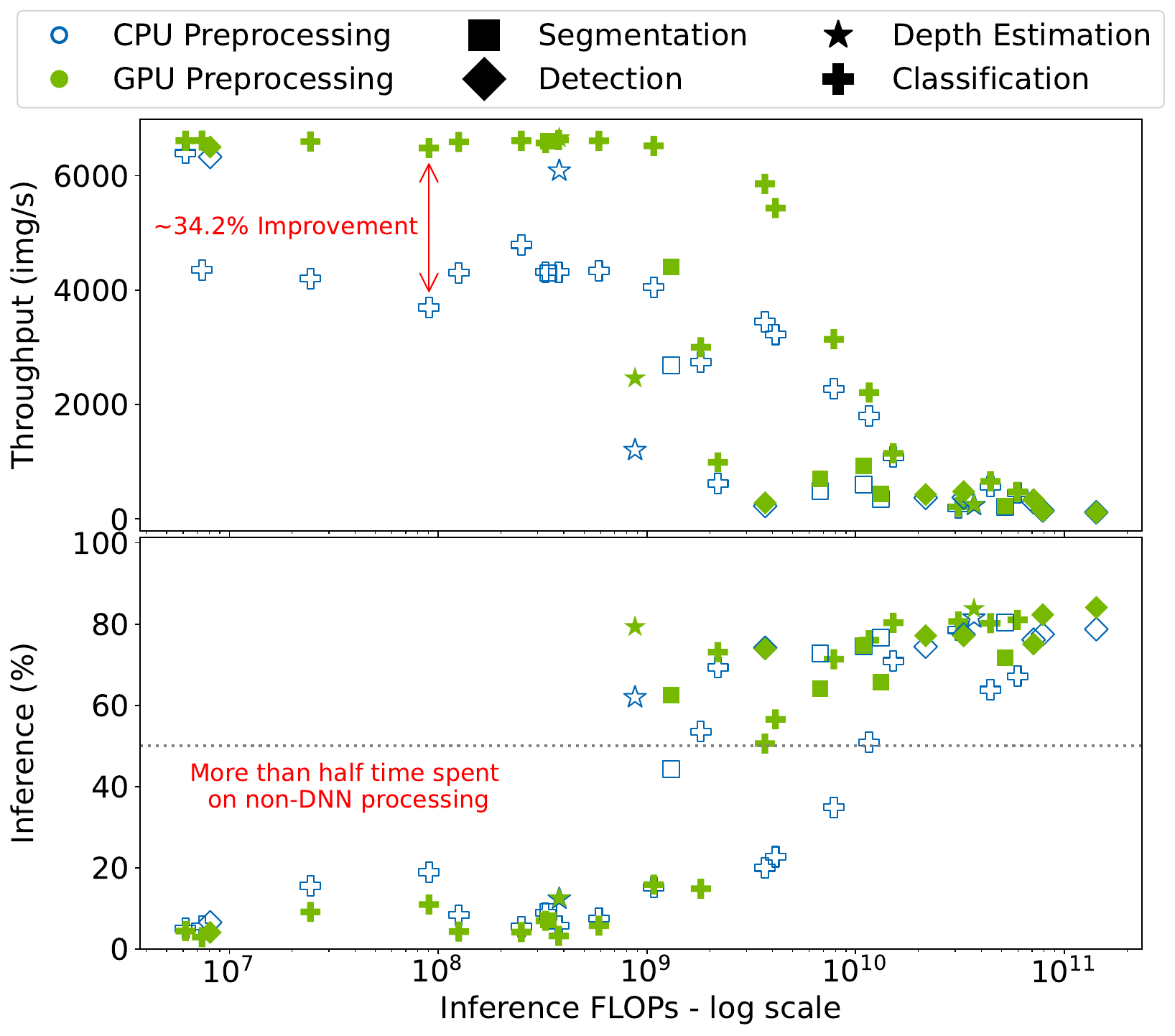}
        \vspace*{-8mm}
    \caption{Throughput and inference time percentage for various HuggingFace models for both CPU/GPU preprocessing.}
    \vspace*{-3mm}

    \label{fig:various_models_analysis}
\end{figure}

% =============================
\vspace{-.3cm}
\section{Related Work}

MLPerf~\cite{reddi2020mlperf} has been a community-driven effort to standardize the benchmarking for AI workloads on different hardware platforms.
Other notable efforts include Alibaba's AI Matrix~\cite{aimatrix}, Fathom~\cite{adolf2016fathom} and DAWNBench~\cite{coleman2017dawnbench}.
Benchmarking end-to-end performance in this way enables a fair comparison of complete AI deployment solutions, including both the hardware system and the software stack.
However, it does not expose specific performance bottlenecks within the AI workload, nor does it specifically attempt to alleviate these bottlenecks through optimized implementations.
To address this shortcoming, inferBench~\cite{zhang2020inferbench}, focuses on server-side inference and compares different serving frameworks including Tensorflow Serving, ONNX runtime, and TrIS, and different \textit{serving formats} such as ONNX, TorchScript, and TensorRT.
Another work, iBench~\cite{brewer2020ibench,brewer2020inference}, focuses on evaluating client-side preprocessing and server-side inference using a custom Flask-based server.
We build upon the findings from these prior benchmarking efforts but focus specifically in identifying the performance bottlenecks within state-of-the-art DNN servers for a number of AI computer vision applications.
Finally, the term ``AI Tax" was coined by Richins et al.~\cite{richins2021ai}, where they studied an AI image processing pipeline with face detection followed by face identification with an Apache Kafka broker in between the two stages to manage data communication. 
Only CPU-based inference was studied, and they found that the time spent in performing DNN inference amounts to only $60\%$, while $35.9\%$ of the latency time is spent in the Kafka broker. 
We improve upon this work through two alternative implementations: An in-memory message broker (Redis), and by investigating the limitations of a fused implementation in Section~\ref{sec:multi-dnn}.
\begin{figure*}[t]
    \centering
    \begin{subfigure}[b]{0.45\linewidth}
        \includegraphics[width=\linewidth]{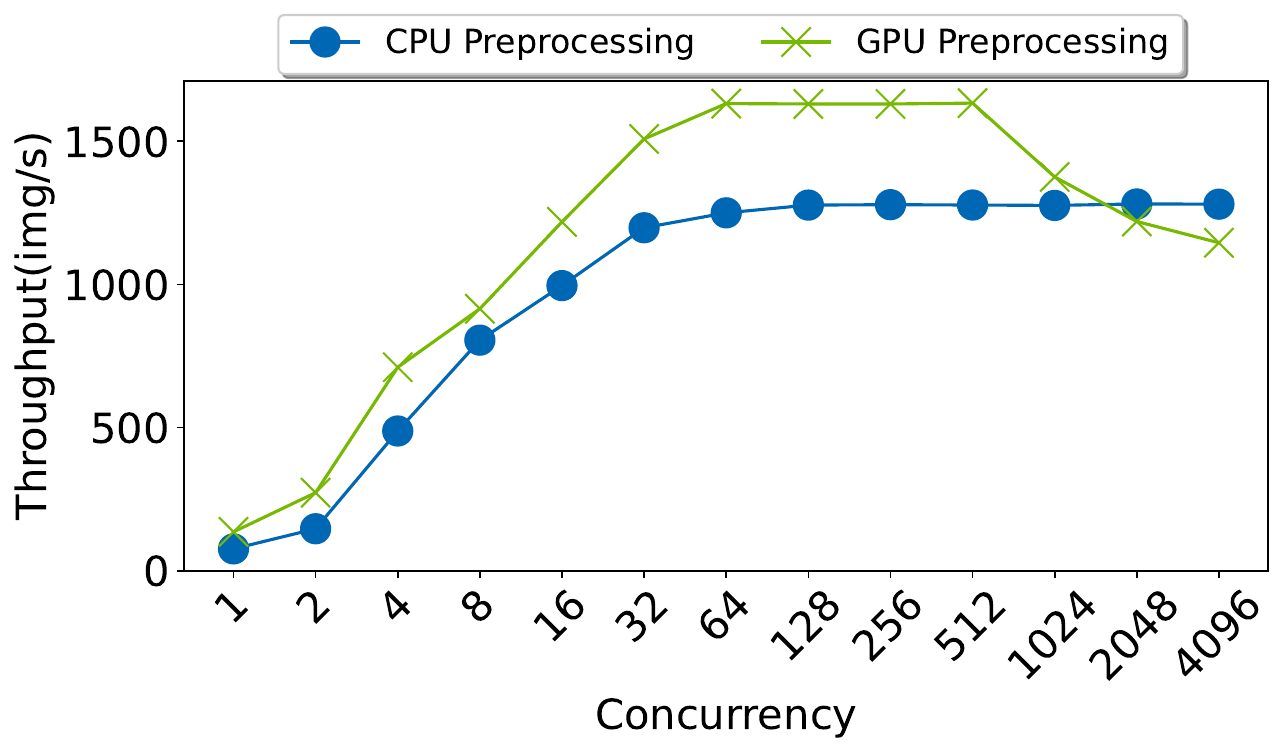}
        \label{fig:concurrency_effect_best_performers_throughput}
    \end{subfigure}
    \hfill
    \begin{subfigure}[b]{0.45\linewidth}
        \includegraphics[width=\linewidth]{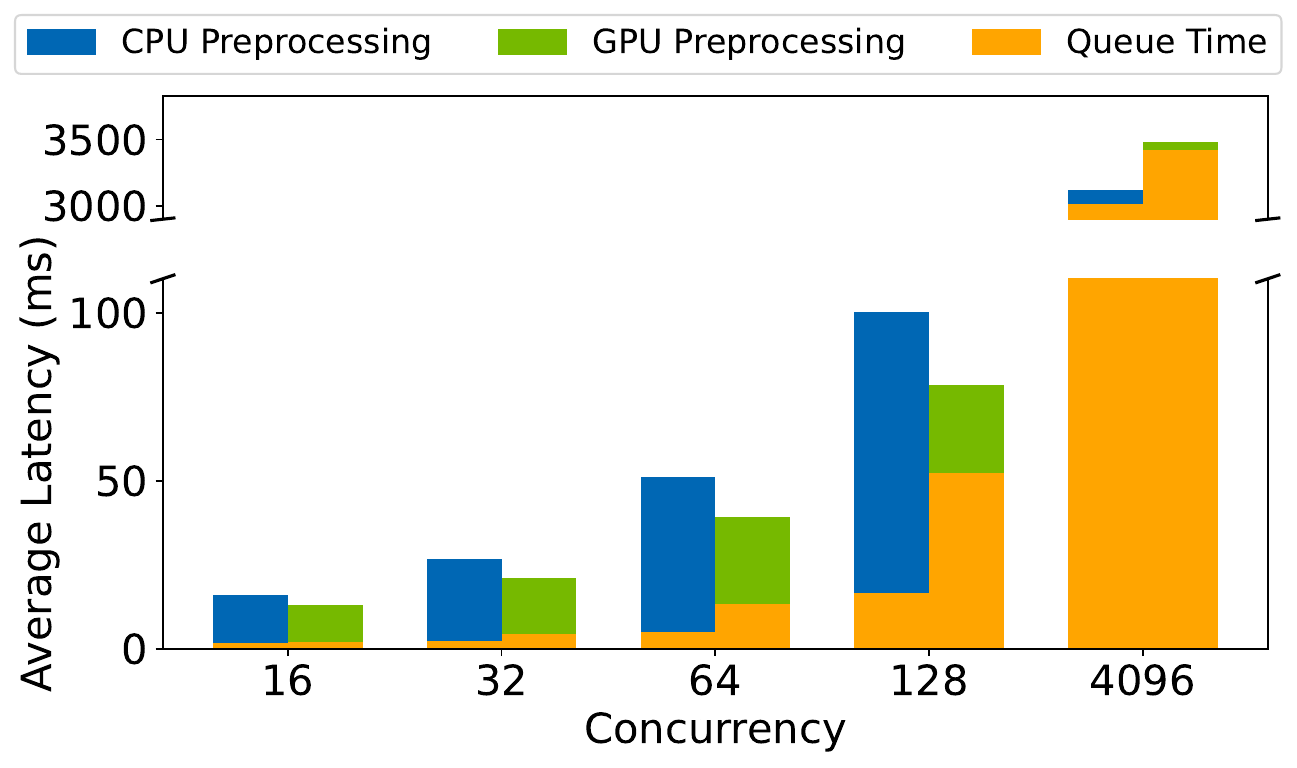}
\label{fig:concurrency_effect_best_performers_latency}
    \end{subfigure}
    \vspace{-.7cm}
    \caption{Throughput, average latency, and queuing time of a throughput-optimized inference server at different concurrencies.}
    \label{fig:concurrency_effect_best_performers}
\end{figure*}
\begin{figure}[!t]
    \centering
    \includegraphics[width=.95\linewidth]{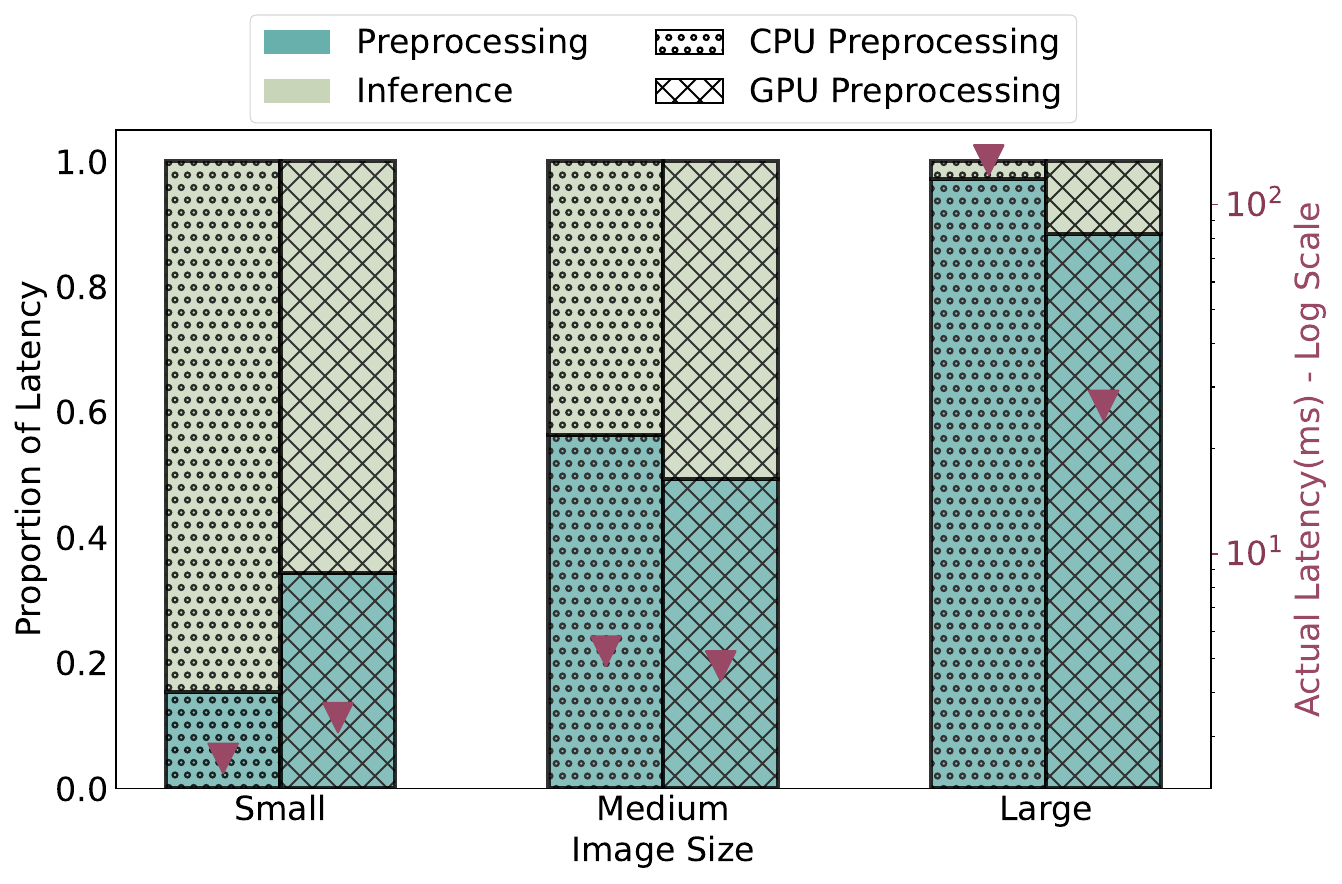}
    \vspace*{-4mm}
    \caption{Latency breakdown of ViT with JPEG preprocessing under zero-load running on TrIS for different image sizes on CPU and GPU preprocessing.}
    \label{fig:latency_breakdown_zero_load}
\end{figure}
%

%==========================================================
\section{Results \& Discussion}

Our goal is to better understand the overheads of DNN serving.
This section presents our results in multiple important settings, spanning across different computer vision DNNs, different hardware setups, different server configurations, and the use of different message brokers---a setting in which we demonstrate considerable improvements compared to recent work~\cite{richins2021ai}.
In all experiments, we use throughput-optimized configurations with TrIS+TensorRT to model production servers as we argued in Section~\ref{sec:software_impact}.
In all cases, DNN inference is performed on the GPU, but we explore both CPU and GPU preprocessing throughout our experiments.
In the following experiments, the preprocessing pipeline consists of JPEG decoding followed by image resizing and normalization.

\subsection{Broad Analysis of Computer Vision DNNs}
\label{sec:various_models_analysis}

To begin our analysis of server overheads, we profile a large number of computer vision DNNs from HuggingFace in Fig.~\ref{fig:various_models_analysis}.
A natural consequence of increasing DNN FLOPs, is that throughput decreases as shown in Fig.~\ref{fig:various_models_analysis} (top).
By benchmarking these models with both CPU and GPU preprocessing, we quantify the improvement from GPU preprocessing to range between -2.9\% to 104\% with an average of 34\% across our models.
As the inference FLOPs increase, the percentage of time taken in inference tends to increase as well.
Fig.~\ref{fig:various_models_analysis} (bottom) quantifies the average time spent on DNN inference from the point at which an image enters the host CPU, until the DNN result is returned to the host CPU.
The remainder of the time is spent on preprocessing, queueing, data transfer, and postprocessing.
Fig.~\ref{fig:various_models_analysis} shows that these DNN overheads dominate inference requests for most models smaller than 5~GFLOPs---efficient image classification DNNs such as ResNet-50 are dominated by non-inference time.
Even for models larger than 10~GFLOPs, 16--49\% of the latency goes towards non-DNN functions, highlighting the importance of further analyzing these overheads.

\subsection{Preprocessing Overhead}

Fig~\ref{fig:latency_breakdown_zero_load} plots the latency and its breakdown into preprocessing and DNN (ViT~\cite{dosovitskiy2020image}) inference under \textit{zero-load} conditions.
We plot latency for both CPU and GPU preprocessing for three representative image sizes\footnote{Small: 4kB 60$\times$70, Medium: 121kB 500$\times$375, Large: 9528kB 3564$\times$2880.} from the ImageNet dataset.
Larger images consume more time in the preprocessing stage as decompression and resizing become more expensive, however, DNN inference is always performed on a standard 224$\times$224 image.
This mimics a realistic server scenario that accepts images from many clients and different resolutions/sizes, but needs to resize them all to a standard size accepted by the DNN.
Interestingly, CPU preprocessing outperforms GPU in terms of latency for small images.
This is likely because the GPU is vastly underutilized in this case.
However, as the image size increases, GPU latency shows marked improvement, becoming significantly better for very large image sizes. 
A clear trend is that the portion of time spent on preprocessing increases with the image size, reaching up to \textbf{56\%, 49\%} in the medium image and up to \textbf{97\%, 88\%} in the large image in cases of CPU and GPU preprocessing respectively.
This analysis demonstrates the importance of explicitly considering the preprocessing functions when designing the hardware for a datacenter server.
Even with accelerated GPU preprocessing, and image sizes from within ImageNet, preprocessing dominates overall latency.

\vspace{-0.1cm}

\subsection{Queuing and Concurrency}

Servers are commonly subjected to high request loads with an objective to maximize throughput while maintaining an acceptable tail latency. 
To build an optimized server, we assume that each node is running at capacity (i.e: receiving a specific number of concurrent requests), and additional requests are routed to other server nodes.
Our goal is thus to maximize the throughput of each node to subsequently minimize the number of nodes required for the whole system.
To assess performance scaling with concurrency in one node of a throughput-optimized system, we test our node under different concurrencies, and we record throughput and average latency. 
The results are illustrated in Fig~\ref{fig:concurrency_effect_best_performers}.
As concurrency increases, throughput increases but latency increases as well. 
GPU preprocessing generally provides higher throughput and lower latency than CPU preprocessing. 
However, GPU preprocessing exhibits a performance decline at very high concurrency, whereas CPU preprocessing saturates, maintaining its output rates under high load.
We postulate that the decline in GPU preprocessing performance at higher concurrency levels stems from GPU memory capacity limitations. 
As the GPU memory saturates, preprocessed inputs queued for inference get temporarily ousted from the GPU memory, necessitating a subsequent reload---a process that incurs additional latency. 
Conversely, CPU preprocessing benefits from a larger main memory that can buffer images until they can be consumed by the GPU.
Critically, Fig.~\ref{fig:concurrency_effect_best_performers} (right) shows that queuing consumes an increasing portion of round-trip latency as concurrency increases, and up to 3 seconds at 4096 concurrency.
However, the optimal concurrencies in this case fall between 64 and 512 where queuing accounts for 34-91\% of the latency.
Even though GPU preprocessing enables higher throughput and lower latencies, more time is spent queuing in the GPU preprocessing case due to resource contention.

\subsection{Throughput Bottlenecks}
\label{sec:throughput_each_stage_images}

\begin{figure}[!t]
    \centering
    \includegraphics[width=1\linewidth]{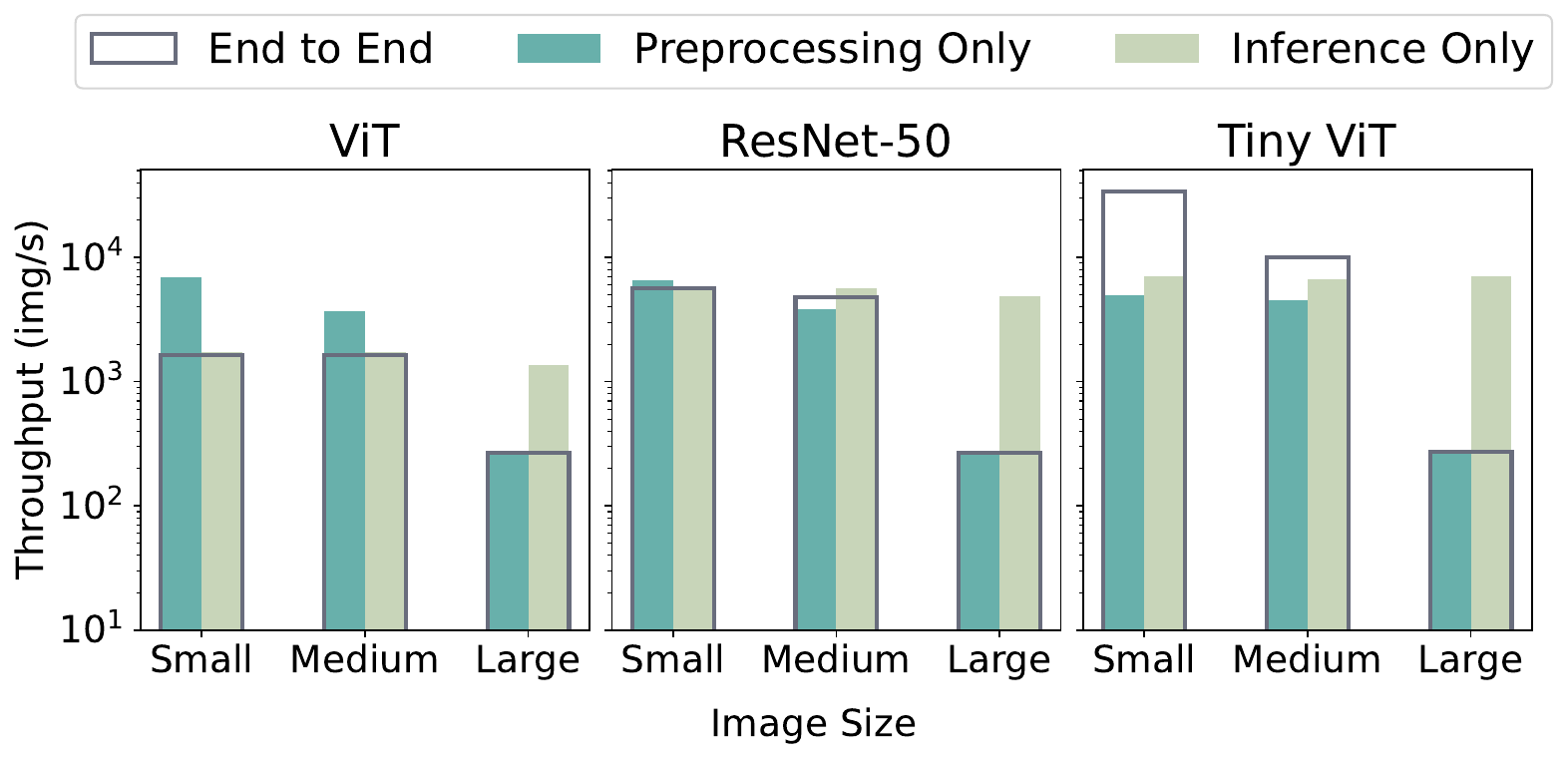}
    \vspace*{-8mm}
    \caption{Comparative throughput analysis of an end-to-end inference server under different models and image sizes with GPU preprocessing.}
    % \vspace{-10px}
\label{fig:image_classification_model_prep_vs_inf_vs_e2e_gpu}
\end{figure}

To understand the impact of preprocessing on throughput, we measure the throughput of GPU preprocessing and inference individually in Fig.~\ref{fig:image_classification_model_prep_vs_inf_vs_e2e_gpu}.
In many cases, the end-to-end throughput is aligned with either the preprocessing or inference stage, indicating the presence of a performance bottleneck in one or the other. 
For ViT-base, a larger model, the inference stage is often the performance bottleneck. 
However, with larger images, preprocessing emerges as the limiting factor, where the throughput of the end-to-end system is just 19.5\% of what's achievable with ViT inference alone. 
The same trend is observed in the smaller ResNet-50 and TinyViT models, confirming the tangible effect that image preprocessing can have on DNN serving throughput.
For medium-sized images, both inference and preprocessing can individually achieve similar throughput, indicating that both need to be optimized to be able to improve end-to-end server throughput.
This highlights the importance of considering all parts of the processing pipeline to achieve some speedup. 
With the increase in performance of current deep learning accelerators, and GPUs in particular, it is clear that there are diminishing returns from simply optimizing deep learning performance on these systems.
Instead, more holistic optimization is now needed, to include additionally heavy preprocessing tasks, especially for computer vision.

An outlier in our results occurs in the case of small/medium images and TinyViT: end-to-end system performance is \textit{faster} than inference-only performance.
We root-caused this issue to data transfer overheads.
Particularly, we only transfer the compressed image to the GPU in the case of end-to-end preprocessing+inference. However, in the inference-only case, we transfer the decoded raw image which is $\sim$5$\times$ larger. 
Even for very fast preprocessing and inference, data transfer may start to limit overall system performance.
Many DNN users are beginning to use such optimized models server-side to save on costs, and our results highlight potential throughput bottlenecks in these cases. 

% Models: 361 to 481

\begin{figure}[!t]
    \centering    
    \includegraphics[width=1\linewidth]{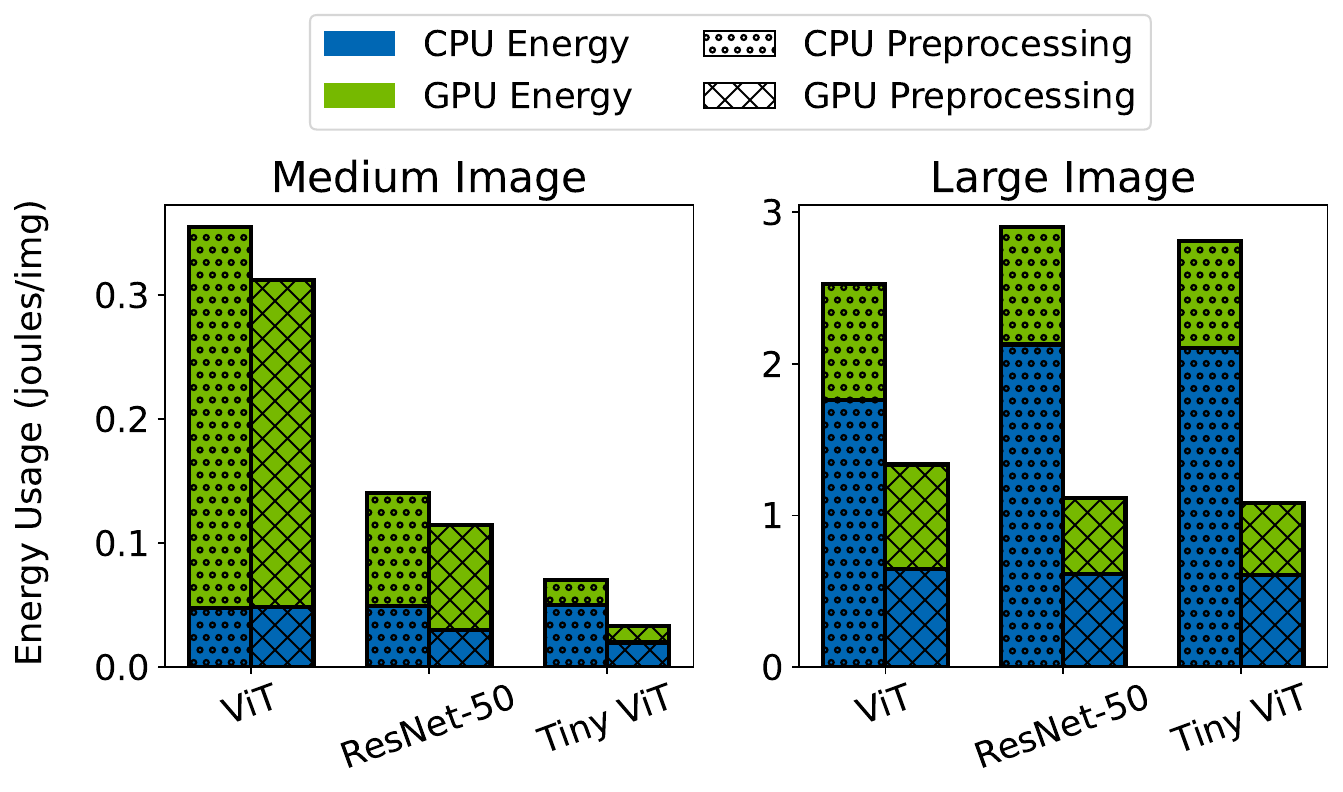}
    \vspace*{-8mm}
    \caption{Energy, measured in joules, expended by the CPU and the GPU per image processed. For each model, left bar is CPU preprocessing and right bar is GPU preprocessing.}
    \vspace*{-3mm}
\label{fig:image_classification_model_best_performing_split_energy_breakdown}
\end{figure}

\subsection{Energy Utilization}

Fig~\ref{fig:image_classification_model_best_performing_split_energy_breakdown} plots the energy utilization per image in different scenarios.
In general, CPU-based preprocessing results in higher energy usage across the board.
This is likely because of the lower device utilization and increased data transfers and memory accesses when the CPU is used for preprocessing and the GPU is used for inference.
When moving from the medium image to the large image, we see a clear increase in CPU energy utilization.
The reason is obvious in the case of CPU preprocessing: a larger image requires more computing time and energy.
For the case of GPU preprocessing, more CPU memory accesses and PCIe transfers are needed to send the larger image to the GPU, resulting in increased energy consumption.
Comparing the GPU portion of energy utilization, it is consistently smaller when the GPU is doing both preprocessing and inference even though it is doing more work in this case.
However, the improved device utilization over-compensates for the additional work, resulting in an overall decrease in average energy per image. 
%\m{double-check energy measurements - I think the current explanation could make sense but it's worthwhile to make sure that everything is correct here.}
%
%As observed, offloading part of the preprocessing to the GPU decreases the CPU's per-image energy consumption. Interestingly, offloading to the GPU always leads a lower per-image energy consumption in the GPU although it is doing more work but that work leads to a higher throughput as well. Given that, the overall energy efficiency is better when preprocessing isn't fully done by the CPU.
% For smaller images, the energy consumption is pretty similar.

\begin{figure}[!t]
    \centering    
    \includegraphics[width=1\linewidth]{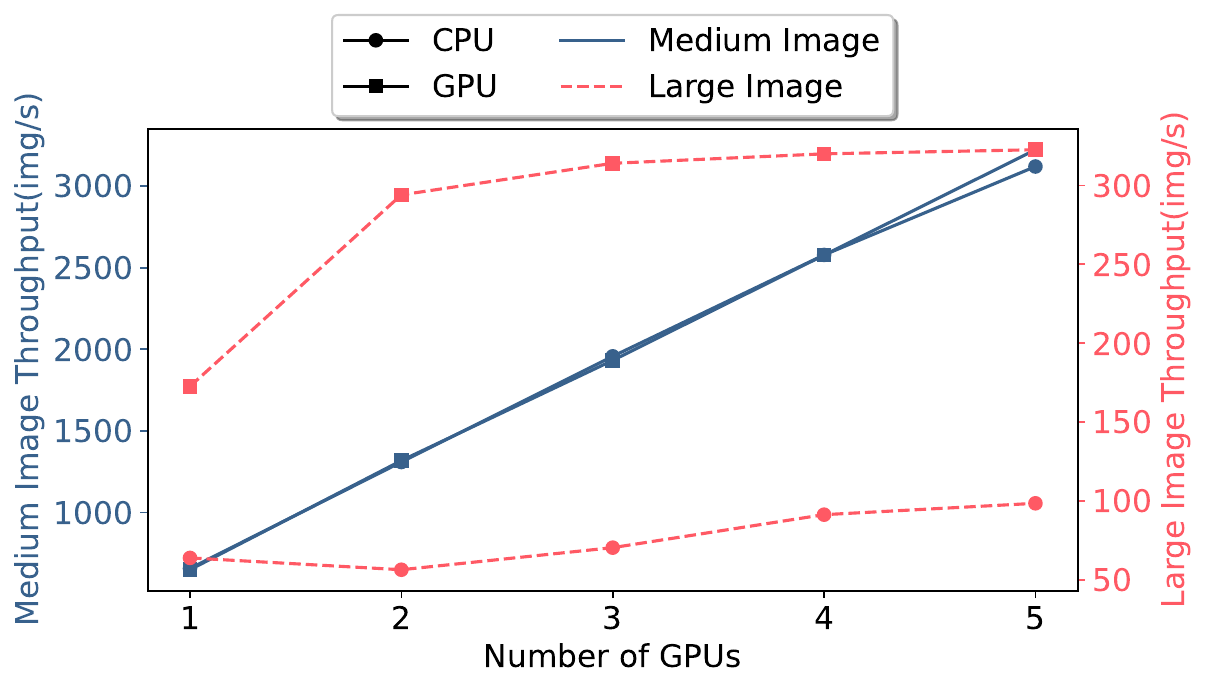}
    \vspace*{-8mm}
    \caption{Throughput variation as we increase the number of GPUs for the medium and large images on different hardware used for preprocessing.}
        \vspace*{-2mm}
    \label{fig:image_classification_vit_multi_gpu}
\end{figure}

\subsection{Multi-GPU Scaling}
\label{sec:multi_gpu}

%The inclusion of multiple GPUs in the system can dramatically transform its performance. In this section, we study such a system and its effect on the throughput. For this study, a separate system equipped with an Intel(R) Xeon(R) Gold 5218 CPU @ 2.30GHz and 5 NVIDIA GeForce RTX 3090 GPUs is utilized. Through the Model Analyzer, the peak throughput is quantified under each system configuration. The results are shown in Figure~\ref{fig:image_classification_vit_multi_gpu}.

To scale performance, it is common to include multiple accelerators connected to each host CPU in a server node.
We study the scaling of ViT-base inference with multiple GPUs in Fig.~\ref{fig:image_classification_vit_multi_gpu}.
Throughput for our medium image exhibits a linear scaling with more GPUs---this happens for both CPU and GPU preprocessing. 
%It is essential to recall, as deduced from Figure~\ref{fig:image_classification_model_prep_vs_inf_vs_e2e_cpu_vs_gpu}, that at this image size, the inference stage acted as the primary performance bottleneck.
%
However, for larger image size, where preprocessing is the performance bottleneck, the increase in GPU count doesn't always translate to an increase in throughput. 
In case of GPU preprocessing, transitioning from a single GPU to dual GPUs introduces a notable throughput enhancement. 
However, further GPU additions result in marginal gains, exposing an underlying performance bottleneck in preprocessing. 
Running inference only shows linear scaling pattern which confirms that inference is not the bottleneck.
%This bottleneck is potentially the JPEG decoding part, where parts of it are still executed on the CPU, which impedes the effective utilization of the additional GPUs. 
When using the CPU for preprocessing, there is minimal change in performance as we increase the number of GPUs since performing the preprocessing consumes a majority of the time and CPU processing cycles, therefore the additional GPUs are wasted, waiting for incoming inference requests from the CPU.

\comment{
\subsection{Video Action Recognition}
\label{sec:video_action_recognition}

In this section, we analyze the maximum achievable throughput in video applications using a similar process to what was used in images in Section~\ref{sec:throughput_each_stage_images}. We try different number of threads for preprocessing to make sure we're finding the optimal system. Current DALI limitations prevent video input batching so maximum batch size in the preprocessing stage is constant at 1. VideoMAE~\cite{tong2022videomae} is used for inference. It is the most downloaded model on HuggingFace~\cite{Hugging_Face_video_2023}. The results are shown in Fig~\ref{fig:video_classification_model_prep_vs_inf_vs_e2e_cpu_vs_gpu}.

\begin{figure}[!h]
    \centering    
    \includegraphics[width=1\linewidth]{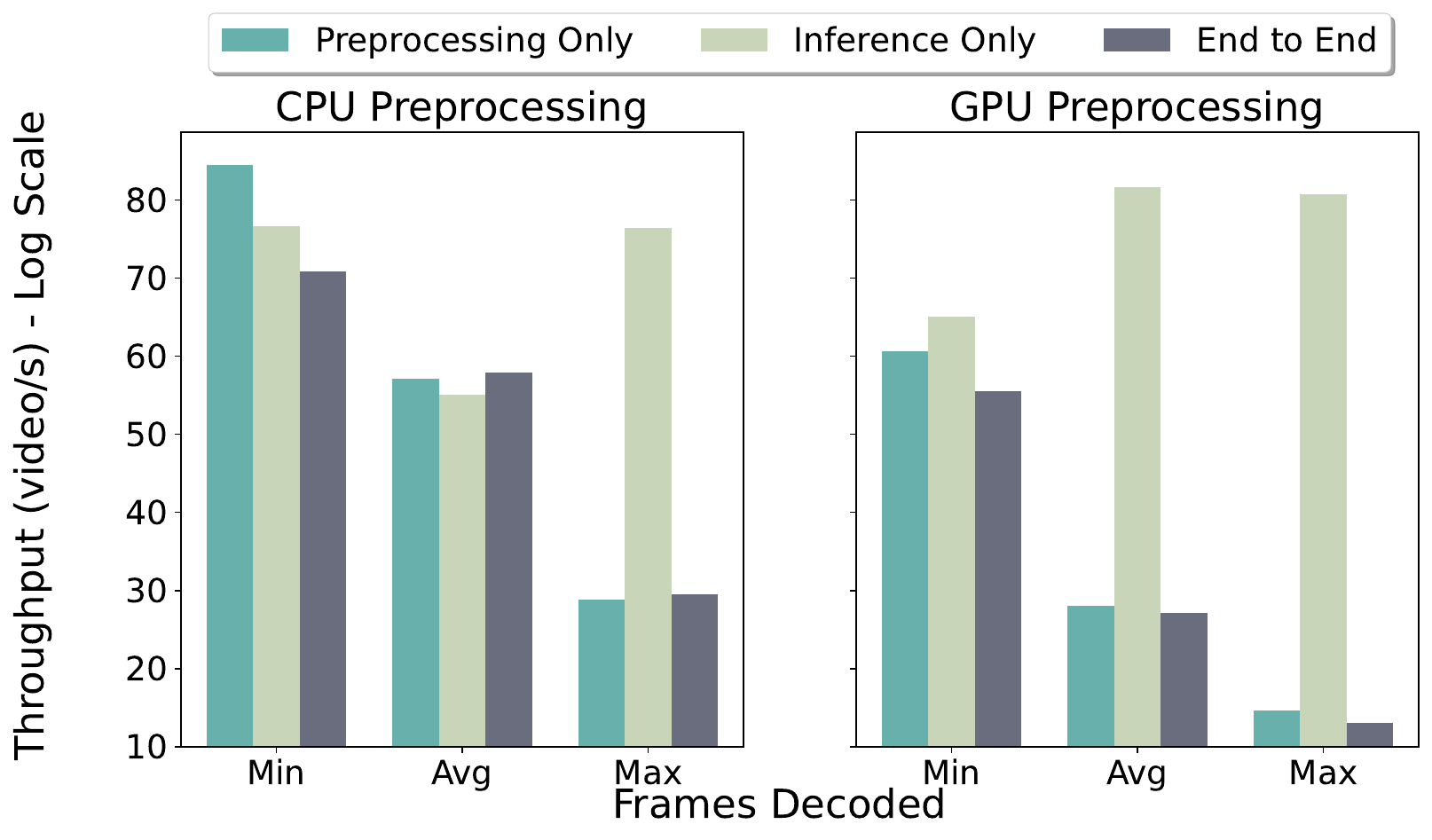}
        \vspace*{-8mm}
    \caption{Action Recognition using VideoMAE - The variation in maximum achievable throughput between performing preprocessing only, inference only and end-to-end inference on TrIS. The left side represents CPU-based preprocessing while the right represents GPU-based processing, with inference consistently GPU-based.}
    \vspace*{-3mm}

    \label{fig:video_classification_model_prep_vs_inf_vs_e2e_cpu_vs_gpu}
\end{figure}
% Models: 211 to 258

A distinct contrast to image processing scenarios is evident. In image processing, the CPU and GPU exhibited comparable performance metrics for smaller images. However, as the image size expanded, the GPU's throughput outperforms the CPU. This behavior is not mirrored in the context of video processing. Remarkably, in all scenarios, the CPU's throughput surpassed that of the GPU. This pattern persisted even as the count of decoded frames increased.
Another observation is that preprocessing is the performance bottleneck in most of the cases. Any trials to optimize the inference here without being accompanied with an optimization in the preprocessing side will likely not lead to any meaningful gains.

% The observed behavior in video action recognition shows the intricate nature of performance dynamics across different data types (images vs. videos) and the pivotal role of the chosen hardware (CPU vs. GPU). It also underscores the potential need to reconsider prevalent assumptions regarding GPU superiority in all deep learning tasks. In the realm of video processing, there might be inherent characteristics or processes that render CPUs more efficient, necessitating further research to pinpoint these factors.
}

\begin{daccomment}    
\subsection{Reproducibility on other hardware}

Given the variability of the hardware available for deep learning tasks nowadays, we run one of the experiments on other systems with different hardware to assess whether the results we see are universal to all systems or they depend on different systems might get different results. we perform similar experiments to the ones that were used in Section~\ref{sec:video_action_recognition}. We chose number of preprocessing threads = 4 as this number previously demonstrated superior throughput. We acknowledge the presence of other concurrent processes, as the systems weren't exclusively dedicated to our experiment, which might introduce some measurement noise. Results are shown in Fig~\ref{fig:videomae_devices_stats}. \ahmed{shall we mention that we're using best performance configurations for each setup alone or repeat the experiments in a similar way to the previous?}

\begin{figure}[!h]
    \centering    
    \includegraphics[width=1\linewidth]{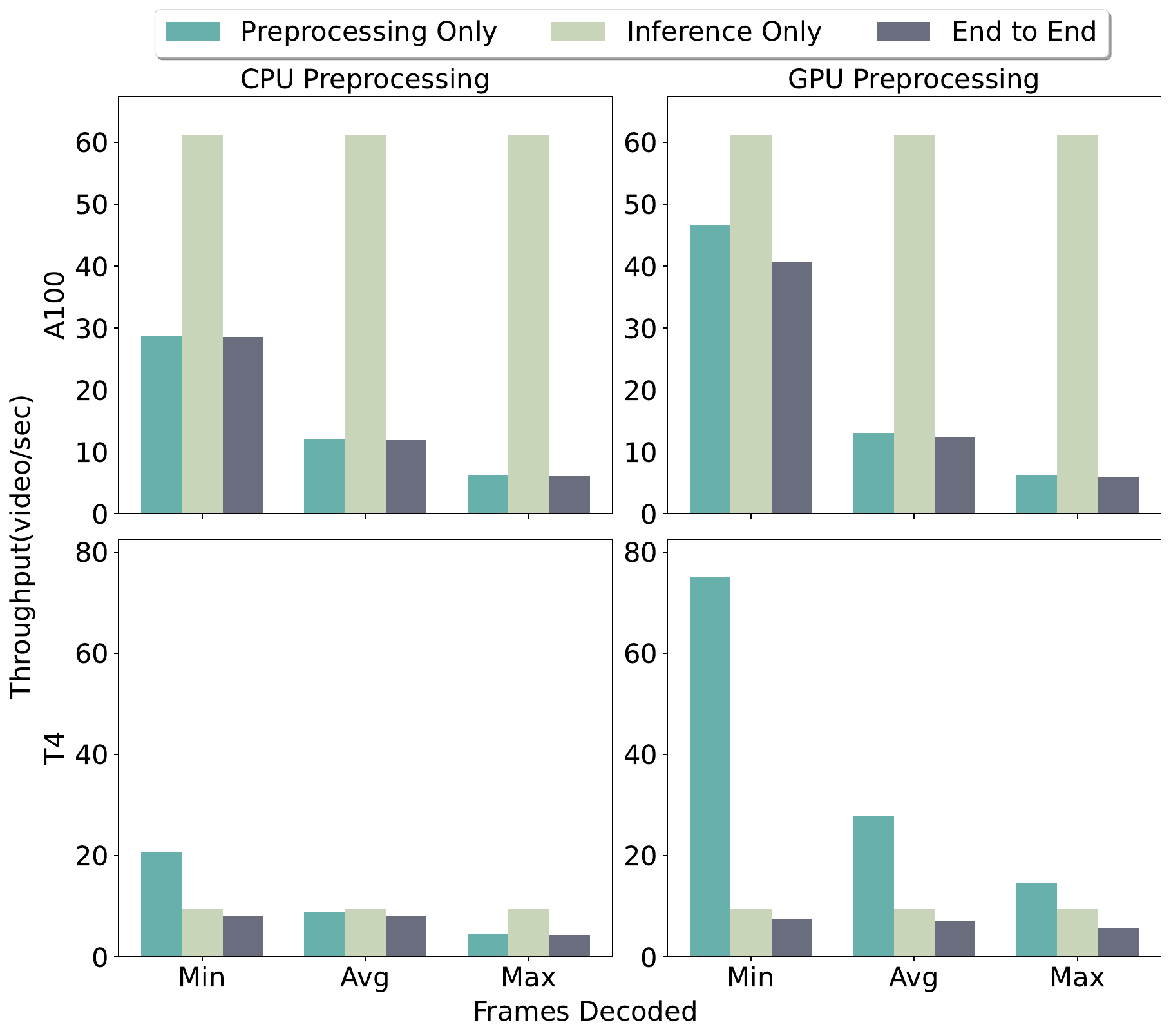}
    \vspace*{-8mm}
    \caption{Action Recognition using VideoMAE running on multiple different systems with different capabilities. Configurations optimal for throughput are plotted.\moh{suggestion: label the y axes with system1 and system2 then have a small table below the figure listing both CPU and GPU?}\ahmed{I made the GPU names friendlier. Not sure about the CPU info. In the text I acknowledge that there might be other other processes running on the system so I feel the CPU info is irrelevant and the main thing I want to highlight is strong GPU vs Weak GPU.}}
    \vspace*{-4mm}
    \label{fig:videomae_devices_stats}
\end{figure}

It is clear that the hardware and system configurations have an effect. In some environments, with less powerful GPUs such as the Tesla T4, inference emerged as the primary performance bottleneck across all configurations. Conversely, in other setups, with stronger GPUs, such as the A100, preprocessing is consistently the primary performance bottleneck. Our findings underscore the dire need for a benchmarking solution that doesn't merely focus on Deep Learning accelerators but instead offers a comprehensive view by measuring the holistic system's performance.

\end{daccomment}

\subsection{Message Brokers in Multi-DNN Systems}
\label{sec:multi-dnn}

\begin{figure}[!t]
    \centering    
    \includegraphics[width=1\linewidth, trim=0 .8cm 0 0]{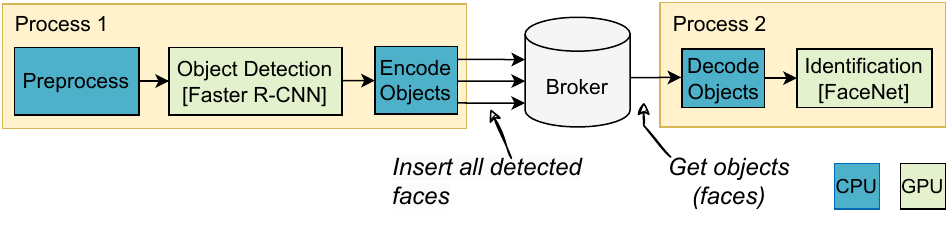}
    % \vspace*{-8mm}
    \caption{A multi-DNN system for face identification.}
    \vspace*{-3mm}
    \label{fig:multi-nn-system}
\end{figure}

\begin{figure}[!t]
    \centering    
    \includegraphics[width=1\linewidth, trim = 0 0 0 .2cm]{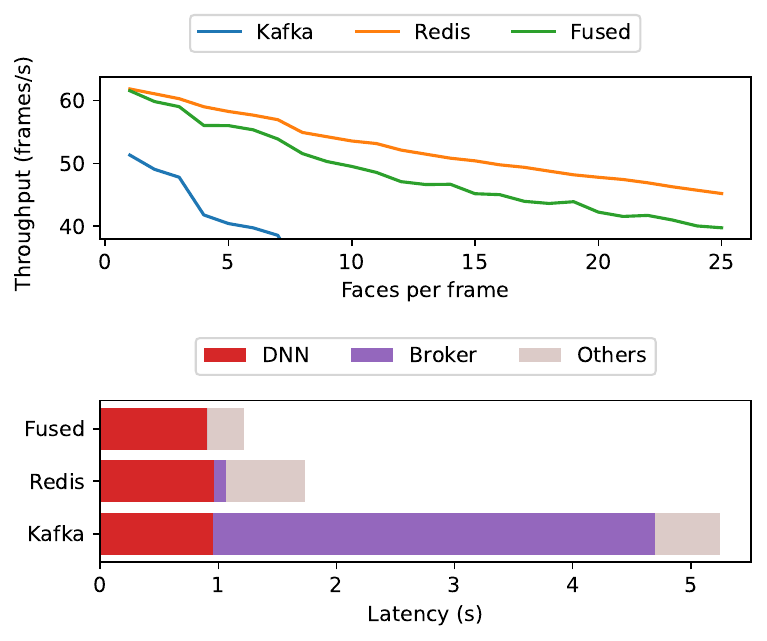}
    \vspace*{-8mm}
    \caption{Throughput and latency breakdown of a multi-dnn systems with different brokers.}
    % \vspace*{-4mm}
    \label{fig:multi-nn-experiment}
\end{figure}

In this section, we analyze a system that contains multiple DNNs connected to each other via a broker similar to the system analyzed in \cite{richins2021ai}, illustrated in Fig~\ref{fig:multi-nn-system}. 
A message broker is useful when two connected processes produce and consume outputs at different rates.
This is the case for the face-detection-then-identification pipeline that we are investigating because one frame can contain multiple faces detected in the first stage using Faster R-CNN, followed by multiple invocations of a face identification DNN in the second stage using FaceNet.
We analyze this pipeline under multiple configurations: Using Apache Kafka as described in prior work~\cite{richins2021ai}, using an in-memory message broker called Redis, and fusing the components of the system into a single process without a message broker in between. 
%The experiments run on the same node and their throughput is measured. Batching is employed in both models at batch size of 32 for R-CNN and batch size of minimum of 128 and number of objects * 32 for FaceNet. Results are shown in Table~\ref{tab:multi-dnn}.
%
Fig~\ref{fig:multi-nn-experiment} shows performance improvement with a memory-backed Redis broker, compared to the disk-backed Apache Kafka, leading to a 125\% improvement in overall system throughput and 67\% improvement in zero-load latency at 25 faces per frame with Kafka taking 71\% and Redis taking 6\% of the total latency. 
It is also observed that the Fused system performs best when the number of detected faces is small.
In this case, the inefficiency of running two stages with different rates is outweighed by the overheads of either message broker.
However, with a high number of detected faces, the gap in throughput between the Fused system and the Redis broker decreases, and eventually Redis outperforms Fused when 9 or more faces are detected.
Our results highlight two key observations. 
First, that message brokers may not always be needed in a multi-DNN system---it depends on the rate mismatch and workload of each DNN.
Second, that in-memory brokers such as Redis significantly outperform disk-based message brokers for multi-DNN systems, thus revising the reported overhead of a face detection system down to just 6\% instead of 36\%~\cite{richins2021ai}.

% \begin{table}[!t]
%     \centering
%     \caption{The effect of different message brokers on the performance of a multi-DNN system.}
%     \begin{tabular}{cccc}\toprule
%   \multirow{2}{*}{\makecell{Number of Detected\\Objects per Image}} & \multicolumn{3}{c}{Throughput(img/s)} \\ \cmidrule{2-4}
%            &  Kafka&Redis& Fused \\ \midrule
%   %1& $58.92_{\pm0.13}$ &$57.88_{\pm0.36}$ & $\mathbf{62.6_{\pm0.08}}$ \\
%   2& $54.33_{\pm0.13}$ &$56.68_{\pm0.26}$ & $\mathbf{60.76_{\pm0.10}}$ \\ 
%   10& $39.95_{\pm0.11}$ &$51.9_{\pm0.20}$ & $\mathbf{52.18_{\pm0.22}}$  \\
%   30& $16.25_{\pm0.07}$ & $\mathbf{36.55_{\pm0.41}}$&$33.84_{\pm0.26}$ \\\bottomrule
%     \end{tabular}
    
%     \label{tab:multi-dnn}
% \end{table}

\comment{
\begin{table}
    \centering
    \caption{Different Brokers effect on performance of a Multi-DNN system. Top 3 use the medium image and bottom 3 use the large image.}
    \begin{tabular}{cccclll}\toprule
  \multirow{2}{*}{\makecell{Number of Detected\\Objects per Image}} & \multicolumn{3}{c}{Throughput(img/s)} &\multicolumn{3}{c}{Latency(s)}\\ \cmidrule{2-4}
           &  Kafka&Redis& Fused & Kafka& Redis&Fused\\ \midrule
  % 1& $58.92_{\pm0.13}$ &$57.88_{\pm0.36}$ & $\mathbf{62.6_{\pm0.08}}$ \\\hline
  2& $54.33_{\pm0.13}$ &$56.68_{\pm0.26}$ & $\mathbf{60.76_{\pm0.10}}$  & 0.78& 0.67&\textbf{0.53}\\ 
  10& $39.95_{\pm0.11}$ &$51.9_{\pm0.20}$ & $\mathbf{52.18_{\pm0.22}}$  & 0.97& 0.87&\textbf{0.62}\\
  30& $16.25_{\pm0.07}$ & $\mathbf{36.55_{\pm0.41}}$ &$33.84_{\pm0.26}$  & 2.04& 4.96&\textbf{0.94}\\\bottomrule
    \end{tabular}
    
    \label{tab:multi-dnn}
\end{table}
}

\vspace{-.3cm}

\section{Conclusion}

\comment{
Our analysis of an optimized inference server for computer vision offers key insights into the performance bottlenecks. Specifically, we observed that the performance bottleneck can manifest either in the preprocessing stage or the inference stage, and this depends on both the size of the inference model and the input size.
Offloading the preprocessing from the CPU to the GPU leads to an increase in the end-to-end performance. However, this optimization is not universally beneficial. Image applications that utilize GPU preprocessing tend to experience performance degradation under high load scenarios, unlike their CPU preprocessing counterparts which maintain consistent performance. Another key implication of using GPU preprocessing is that it consumes a portion of the GPU resources, which could have been allocated to inference tasks. Lastly, a strategy of simply adding more GPUs to the system does not guarantee improved performance. Such an addition proves beneficial only when the inference stage is the performance bottleneck. In scenarios where preprocessing forms the bottleneck, adding more GPUs, even when these GPUs are employed for preprocessing tasks, provides minimal performance enhancement.
Other components in the system can also affect the performance such as brokers in a multi-DNN system. Fusing the components is sometimes better than a complex architecture but if a complex architecture is needed, choosing an efficient broker is key to the performance of the system.
}

In this paper, we benchmarked vision DNNs on a throughput-optimized inference server to analyze system performance, identify performance bottlenecks, and quantify DNN serving overheads.
%
% a broad analysis of vision DNNs
Our broad analysis of vision DNNs clearly demonstrated that inference does not dominate performance on modern GPUs, especially for DNNs less than 5~GFLOPs.
% zero load latency
We proceeded investigate the sources of performance bottlenecks and found that standard preprocessing on common image sizes can account for a large portion (>50\%) of the zero-load DNN serving latency, even when accelerated GPU preprocessing is used.
% queuing accounts for X%
Under high concurrency, we further found that queuing accounted for $\sim$60\% of total latency. 
% because GPUs have gotten so good in DNNs, throughput can be limited by either preprocess or inference
% this is also true for multi-gpu scaling 
From a throughput perspective, we found that modern GPUs have become very efficient in processing DNNs and once again preprocessing could limit system performance, especially for single CPU, multi-GPU systems.
Accelerating preprocessing on the GPU using the NVIDIA DALI library can alleviate these scaling limitations but only to a certain extent due to batched preprocessing.
However, overall performance can still be throttled by preprocessing beyond two GPUs.
% more complex multi-dnn systems can benefit from message brokers for rate matching, but the existence and choice of a broker depends the exact application - it is not always needed.
Finally, we investigated the impact of message brokers between two DNNs and found that prior work has overestimated their overhead because of the reliance on Apache Kafka.
We additionally investigated Redis and a Fused implementation, showing that the broker overhead can be as low as 6\% and we boosted performance by 2.25$\times$ compared to prior work.
Our work provides a clearer understanding of DNN servers for computer vision tasks, and lays the foundations for optimized system design.

\comment{

\section{Summary \& Future Work}

In this study, we analyzed computer vision inference systems, examining both image and video inputs. Our focus was to understand how various server parameters influence system performance. A critical part of our investigation centered on understanding the implications of offloading parts of the preprocessing from the CPU to the GPU, especially concerning performance and energy utilization. Through our research, we identified the performance bottlenecks in an optimized system under varied conditions and delved into its scaling properties.

Our findings have clear implications: while much attention has historically been centered on optimizing inference, our results spotlight the importance of considering other stages, like preprocessing. At times, it's the preprocessing stage that prevents the realization of performance gains.

Looking ahead, several avenues need further exploration. Expanding the analysis beyond computer vision to encompass other applications is one promising direction. Additionally, examining the performance on alternative hardware, such as TPUs and custom accelerators, could yield valuable insights. Another direction is offloading parts of the system that may be the performance bottlenecks, such as the preprocessing stage, to a custom harware or an FPGA.  
Moreover, we recognize a pressing need for a benchmarking system akin to the robustness and acceptance of MLPerf but encompassing end-to-end system performance measurements. \ahmed{shall I remove this line because I removed the reproducibility on other hardware section?}
}

\vspace{-.3cm}

\begin{acks}
This project is supported in part by Intel Corporation funding. We would like to thank Mahesh Iyer, Nilesh Jain, Aravind Dasu, Gregg Baeckler, and Ilya Ganusov for insightful discussions and feedback.
\end{acks}

\bibliographystyle{ACM-Reference-Format}
\bibliography{sample-base}

\appendix

\end{document}